\documentclass[12pt]{iopart}
\usepackage{amssymb}
\usepackage[dvips]{graphicx}
\usepackage{hyperref}
\hypersetup{colorlinks=blue}

\begin{document}

\title[Formation of Intermediate Coupling Optical Polarons and Bipolarons ...]{Formation of Intermediate Coupling Optical Polarons and Bipolarons in Two-Dimensional Systems}

\author{S.Dzhumanov$^{1}$, P.J.Baymatov$^{2}$, N.P.Baymatova$^{2}$, Sh.T.Inoyatov$^{2}$ and O.Ahmedov$^{2}$}

\address{$^1$Institute of Nuclear Physics, Uzbek Academy of Science, Tashkent, 100214, Uzbekistan\\
$^{2}$Namangan State University, Department of Physics, 716003
Namangan, Uzbekistan}
\ead{dzhumanov@rambler.ru}
\begin{abstract}
The formation of the optical polaron and bipolaron in
two-dimensional (2D) systems are studied in the intermediate
electron-phonon coupling regime. The total energies of 2D polaron
and bipolaron are calculated by using the Buimistrov-Pekar method of
canonical transformations and analyzed in the weak, intermediate and
strong coupling regimes. It is shown that the electron-phonon
correlation significantly reduces the total energy of 2D polaron in
comparison with the energy of the strong-coupling (adiabatic)
polaron. A charge carrier in polar crystals remains localized in a
2D potential well when the electron-phonon coupling constant
$\alpha$ is greater than the critical value $\alpha_{c}\simeq2.94$,
which is much lower than a critical value of the electron-phonon
coupling constant $\alpha$ for a 3D system. The critical values of
the electron-phonon coupling constant $\alpha$ and the parameter of
Coulomb repulsion between two carriers
$\beta=1/(1-\varepsilon_{\infty}/\varepsilon_{0})$ (where
$\varepsilon_{\infty}$ and $\varepsilon_{0}$ are the high frequency
and static dielectric constants, respectively), which determine the
bipolaron stability region, are numerically calculated. The obtained
results are compared with the ones obtained by using the Feynman
path integral method and the modified Lee-Low-Pines unitary
transformation method.
\end{abstract}

\vskip2pc

 \maketitle
\section{Introduction}
As is well known, charge carriers (electrons and holes) is ionic
crystals interact with polar optical phonons and they are
self-trapped at their sufficiently strong coupling to the phonons
with the formation of optical polarons (called as the Pekar or
Fr\"{o}hlich polarons) \cite{1,2,3}. One distinguishes three
distinct regimes of electron-phonon coupling \cite{2}: (i) the
weak-coupling regime describes the correlated motions of the lattice
atoms and the quasi-free charge carriers which remain in their
initial extended state, (ii) the intermediate-coupling regime
characterizes the self-trapping of a charge carrier which is bound
within a potential well produced by the polarization of the lattice
in the presence of the carrier and follows the atomic motions, and
(iii) the strong-coupling regime describes the other condition of
self-trapping under which the lattice atoms no longer follow the
charge carrier motion and the self-trapping of carriers is usually
treated within the adiabatic approximation (i.e. lattice atoms
remain at their fixed positions). Under certain conditions, two
charge carriers interacting with the lattice vibrations and with
each other can form a bound state of two carriers in polar materials
within a common self-trapping well. Since the attractive interaction
of the electron-phonon coupling in these systems is strong enough to
overcome the Coulomb repulsion between two carriers. The
self-trapped state of the pair of charge carriers is termed a
bipolaron. In the last few decades, the bipolaron problem has been a
focus of attention due to its importance in semiconductor technology
and in the bipolaronic mechanism of superconductivity (see Refs.
\cite{4,5,6,7,8,9,10,11,12}). After the discovery of the layered
high-$T_{c}$ cuprate superconductors, the study of bipolarons has
attracted the revived interest because some researchers believe that
the bipolaron is one of the possible candidate for explaining
high-$T_c$ superconductivity. The mechanism of high-$T_c$
superconductivity of large bipolarons was proposed by Emin and
Hillary \cite{7}. This mechanism is based on the Bose-Einstein
condensation of bipolarons as discussed by Schafroth \cite{13}.
Another mechanism of high-$T_c$ superconductivity driven by the
superfluid single particle and pair condensation of large bipolarons
and polaron Cooper pairs was proposed in \cite{14,15}. The
possibility of such a novel superconductivity depends on the
existence of polaron and bipolarons in the superconducting
materials.  The above electron-phonon coupling regimes are
characterized by the dimensionless Fr\"{o}hlich coupling constant
$\alpha$. The possible ranges of the values of $\alpha$
characterizing the weak, intermediate and strong coupling regimes
depend on the type of self-trapped charge carriers and the
dimensionality of the system. The polaron and bipolaron ground state
in the strong coupling limit (i.e. in the adiabatic approximation)
have been studied by many authors for 3D and 2D systems
\cite{6,7,8,9,10,11,16,17}. However, with decreasing $\alpha$, it is
necessary to take into account the electron-phonon correlation which
reduces the energies of the polaron and bipolaron in comparison with
those of the adiabatic polaron and bipolaron. Therefore, a
quantitative treatment of the (bi)polaron problem in the
intermediate-coupling regime was also a subject of numerous studies
in the past three decades (see Refs \cite{6,8,9,10,16,18,19}). The
ground state energies of the polaron and bipolaron have been
calculated by using several approximations, such as the Feynman path
integral method \cite{6,9,10,16,20}, the Lee-Low-Pines (LLP) unitary
transformation method \cite{16,18} and the operator formalism
\cite{8,18}. According to different intermediate coupling
treatments, the formation of the polaron and bipolaron becomes
possible only if the values of $\alpha$ are greater than the certain
critical values $\alpha_c$. The previous calculations have shown
that the (bi)polaron is created more easily in 2D systems than in 3D
ones and the stability region for bipolaron formation is much
broader in 2D case as compared with 3D case. So far, calculations of
the ground-state energies of the (bi)polaron, the values of
$\alpha_c$ and the stability region of the bipolaron are not
conclusive. Some of the estimated values of $\alpha_{c}$ differ
greatly in magnitude. Further, the calculation methods used to study
the (bi)polaron problem suffer from the certain drawbacks (see Refs.
\cite{9,18}). In particular, the Feynman path integral method is
more accurate for the calculation of the polaron energy
\cite{9,16,18}, but it fails to describe correctly the bipolaron
formation. Moreover, the functional for the bipolaron energy has
rather a cumbersome form which requires the tedious numerical
calculations. Therefore, it is expedient to use the another method
for the calculation of the polaron and bipolaron energies in the
intermediate-coupling regime. One of such methods is the
Buimistrov-Pekar method \cite{21}. In this approach, the
electron-phonon correlation is taken into account through the
displacement amplitude $F_{\vec{q}}(\vec{r})$ of the form:
\begin{eqnarray}\label{1Eq}
F_{\vec{q}}(\vec{r})=f_{\vec{q}}+g_{\vec{q}}\exp(-i\vec{q}\vec{r}),
\end{eqnarray}
where $f_{\vec{q}}$ and $g_{\vec{q}}$ are the variational
parameters, which are determined from the condition for minima of
the total energy, $\vec{r}$ is the electron coordinate, $\vec{q}$ is
the phonon wave vector.

The second term in Eq. (\ref{1Eq}) takes into account the
electron-phonon correlation effect that leads to the reduction of
the energy of a 3D adiabatic polaron when $\alpha>\alpha_c$. At
$\alpha<\alpha_c$ the present method leads to the delocalized state
of a polaron, and the total energy of the polaron is given by
$E_{p}=\alpha\hbar\omega_{o}$ (where $\omega_{o}$ is the
longitudinal optical (LO) phonon frequency). In the Feynman path
integral variational approach \cite{20}, the delocalization of the
polaron at a finite value of $\alpha$ does not occur and the lowest
polaron energy is obtained. However, at $\alpha>\alpha_c$ the
Buimistrov-Pekar method taking into account an important part of the
electron-phonon correlation gives also reasonable results for the
energy of the polaron. The main advantage with this method is that
it is simple and does not require tedious numerical calculations.
Recently, the Buimistrov-Pekar method was applied to study both the
free bipolaron and the bound bipolaron in 3D systems
\cite{22,23,24}. The existence of polarons and bipolarons in
high-$T_c$ cuprates and other materials has been indicated by
several experiments \cite{25,26,27,28,29}. The $CuO_{2}$-based
layered high-$T_{c}$ materials are believed to be
quasi-two-dimensional systems and in an intermediate electron-phonon
coupling regime. So far, the problems of the 2D polaron and
bipolaron are not studied sufficiently and the possibility of
formation of intermediate-coupling polarons and bipolarons in 2D
systems within the Buimistrov-Pekar formalism is not explored. The
aim of the present paper is to study the formation of the
intermediate coupling optical polaron and bipolaron in 2D systems by
using the Buimistrov-Pekar method and to expose the important
features of this method. We calculate the ground-state energies of
the intermediate coupling 2D (bi)polarons, the critical values of
$\alpha$, the parameter of the Coulomb repulsion between two
carriers $\beta=1/(1-\varepsilon_{\infty}/\varepsilon_{0})$ (where
$\varepsilon_{\infty}$ and $\varepsilon_{0}$ are the high frequency
and static dielectric constants) for the formation of 2D polaron and
bipolaron. We discuss the obtained results and compare them with
previous ones obtained by using the Feynman path integral method and
the modified LLP unitary transformation method \cite{16,30}.
\section{Formation of intermediate-coupling polarons}
The Hamiltonian and variational wave function describing the
interacting system of electron (or hole) and LO phonons can be
written as \cite{22,24}
\begin{eqnarray}\label{2Eq}
H_{p}=-\frac{\hbar^{2}}{2m^{*}}\Delta+\sum_{q}\left[V_{q}b_{q}\exp(iqr)+
V^{*}_{q}b^{+}_{q}\exp(-iqr)\right]+\sum_{q}\hbar\omega_{o}b^{+}_{q}b_{q},
\end{eqnarray}
and
\begin{eqnarray}
\Psi=\Phi_{ph}\varphi(r)=U|0\rangle\varphi(r),
\end{eqnarray}
where
\begin{eqnarray}
U=\exp\left[\sum_{q}\left(F_{q}(r)b^{+}_{q}-F_{q}^{*}(r)b_{q}\right)\right],\quad
U^{*}U=1, \quad \langle0|0\rangle=1,
\end{eqnarray}
\begin{eqnarray}\label{10Eq}
\frac{|V_{q}|^{2}}{\hbar\omega_{o}}=\frac{2\pi\hbar\omega_{o}
l_{o}}{L^{2}q}\alpha, \quad
l_{o}=\sqrt{\frac{\hbar}{2m^{*}\omega_{o}}},\quad
\alpha=\frac{e^{2}}{2\pi\omega_{o}l_{o}}\left(\frac{1}{\varepsilon_{\infty}}-\frac{1}{\varepsilon_{0}}\right),
\end{eqnarray}
$m^*$ is the effective mass of the carrier before polaron formation,
$V_{\vec{q}}$ is the 2D form factor of the electron-phonon
interaction, $L^2$ is the  size (or surface) of the 2D system,
$b^{+}_{q}(b_{q})$ is the creation (annihilation) operator of the LO
phonon with the wave vector $\vec{q}$ and the frequency
$\omega_{o}$, $\varphi(r)$ is the one-electron wave function
$\Phi_{ph}$ represents the phonon part of the wave function,
$|0\rangle$ is the unperturbed zero-phonon state satisfying
$b_{\vec{q}}|0\rangle=0$ and $\langle0|0\rangle=1$. Averaging the
Hamiltonian (\ref{2Eq}) over $\Phi_{ph}$, we have
\begin{eqnarray}\label{3Eq}
\tilde{H}_{p}=-\frac{\hbar^{2}}{2m^{*}}\Delta+\sum_{q}\left[\frac{\hbar^{2}}{2m^{*}}\left|\nabla
F_{q}(r)\right|^{2}+\hbar\omega_{o}\left|F_{q}(r)\right|^{2}+\right.\nonumber\\
\left.V_{q}F_{q}(r)\exp(iqr)+V^{*}_{q}F^{*}_{q}(r)\exp(-iqr)\right]
\end{eqnarray}
After substituting Eq.(\ref{1Eq}) into Eq.(\ref{3Eq}), averaging the
Himiltonian (\ref{3Eq}) over $\varphi(r)$, and minimizing the energy
$E_{p}=\langle\varphi|\tilde{H}_{p}|\varphi\rangle$ with respect to
$f_{\vec{q}}$ and $g_{\vec{q}}$, we obtain the following functional
for the total energy of a 2D polaron
\begin{eqnarray}\label{4Eq}
E_{p}=K-\sum_{q}\frac{|V_{q}|^{2}}{\hbar\omega_{o}}W^{2}_{q}-
\sum_{q}\frac{|V_{q}|^{2}}{\hbar\omega_{o}}\frac{[1-W^{2}_{q}]^{2}}{1-W^{2}_{q}+l_{o}^{2}q^{2}},
\end{eqnarray}
where
%
%\begin{eqnarray}
$$
K=\frac{\hbar^2}{2m^{*}}\int
d^{2}r\left(\frac{\partial\varphi}{\partial r}\right)^{2},
$$
$$
W_{q}=\int
d^{2}r\exp(iqr)\varphi^{2}(r),\nonumber\\
\sum_{\vec{q}}...=\left(\frac{L}{2\pi}\right)^{2}\int d^{2}q...
$$
%\end{eqnarray}
%

The latter sum in Eq. (\ref{4Eq}) is the correction term arising
from the electron-phonon correlation caused by the second term in
Eq.(\ref{1Eq}). The functional (\ref{4Eq}) without this term
($g_{\vec{q}}=0$) determines the energy of the strong-coupling
adiabatic polaron. For simplifying the calculations, we choose the
electron wave function in the form
\begin{eqnarray}
\varphi(r)=N\exp(-\delta^{2}r^{2}),\quad \pi N^{2}=2\delta^{2},\quad
W^{2}_{q}=\exp(-q^{2}/4\delta^{2})
\end{eqnarray}
Then we obtain the following expression for the energy (in units of
$\hbar\omega_{o}$) of the intermediate coupling 2D polaron:
\begin{eqnarray}\label{5Eq}
E_{p}=2\mu^{2}-\sqrt{\pi}\mu\alpha-2\mu\alpha\int\limits_{0}^{\infty}dt
\frac{[1-\exp(-t^{2})]^{2}}{1-\exp(-t^{2})+4\mu^{2}t^{2}},
\end{eqnarray}
where $ \mu=l_{o}\delta.$

In the weak and strong coupling limits, we obtain from Eq.
(\ref{5Eq})
\begin{eqnarray}\label{6Eq}
E_{p}=\left\{\begin{array}{ll}
\vspace{0.4cm}-\frac{\pi}{2}\alpha\\
\vspace{0.4cm} -\frac{\pi}{8}\alpha^{2}
\end{array}
\ \ \begin{array}{ll}
\vspace{0.4cm}\textrm{for}\quad\alpha<\alpha_{c}=2.94\\
\vspace{0.4cm}\textrm{for}\quad\alpha\rightarrow\infty
\end{array}
\right.
\end{eqnarray}

The Feynman path integral method and the LLP unitary transformation
method applied to the polaron problem, give the following
expressions for the energies of the polarons \cite{16}:
\begin{eqnarray}\label{9Eq}
E^{F}_{p}=\frac{(v-w)^{2}}{2v}-\frac{\alpha}{2}\sqrt{\frac{\pi}{2}}
\int\limits_{0}^{\infty}dt\frac{\exp(-t)}{\sqrt{D(t)}},\nonumber\\
D(t)=\frac{w^{2}}{2v^{2}}t+\frac{v^{2}-w^{2}}{2v^{3}}[1-\exp(-vt)]
\end{eqnarray}
and
\begin{eqnarray}\label{Eq011}
E^{LLP}_{p}=\frac{\lambda}{2}-\alpha\int\limits_{0}^{\infty}dt\frac{\exp[-(1-\gamma)^{2}t^{2}/\lambda]}{1+\gamma^{2}t^{2}},
\end{eqnarray}
where $v$, $w$ and $\lambda$, $\gamma$ are the respective
variational parameters.

The energies of the 2D polarons in the weak and strong coupling
limits follow directly from Eqs. (11) and (12):
\begin{eqnarray}\label{00Eq}
E_{p}^{F}=\left\{\begin{array}{ll}
\vspace{0.4cm}-\frac{\pi}{2}\alpha-\frac{\pi^{2}}{216}\alpha^{2}\\
\vspace{0.4cm} -\frac{\pi}{8}\alpha^{2}
\end{array}
\ \ \begin{array}{ll}
\vspace{0.4cm}\textrm{for}\quad\alpha\rightarrow0\\
\vspace{0.4cm}\textrm{for}\quad\alpha\rightarrow\infty
\end{array}
\right.
\end{eqnarray}
\begin{eqnarray}\label{17Eq}
E_{p}^{LLP}=\left\{\begin{array}{ll}
\vspace{0.4cm}\ -\frac{\pi}{2}\alpha\\
\vspace{0.4cm} -\frac{\pi}{8}\alpha^{2}
\end{array}
\ \ \begin{array}{ll}
\vspace{0.4cm} \textrm{for}\quad\alpha<\alpha_{c}=3.62\\
\vspace{0.4cm} \textrm{for}\quad\alpha\rightarrow\infty.
\end{array}
\right.
\end{eqnarray}
These results can be compared with Eqs. (\ref{5Eq}) and (\ref{6Eq}).
As can be seen from Eqs. (\ref{6Eq}), (\ref{00Eq}) and (\ref{17Eq}),
all three methods in the strong-coupling limit give the same result.
For compairing, in Fig.1, the energies of 2D polarons (in units of
$\hbar\omega_{o}$) obtained from the Eqs. (\ref{5Eq}) and
(\ref{9Eq}) are plotted as a function of $\alpha$.
\begin{figure}
\begin{center}
\includegraphics[width=0.6\textwidth]{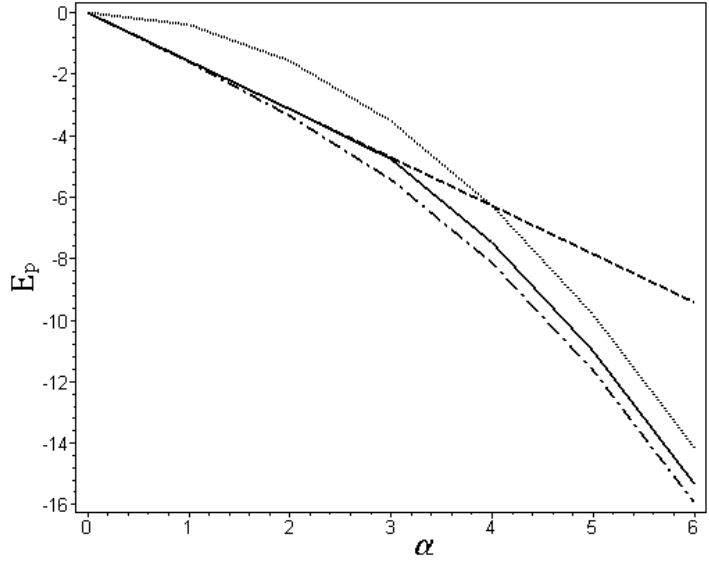}
\caption{\footnotesize The energies of the 2D polarons (in units of
$\hbar\omega_{o})$ as a function of $\alpha$ for the strong coupling
(dotted curve) and weak coupling (dashed line) limits, calculated by
using Eq. (\ref{6Eq}), for the intermediate coupling limit (solid
curve), calculated by using Eq.(\ref{5Eq}). Thin solid curve
calculated by the Feynman path integral method, Eq. (\ref{9Eq}).}
\end{center}
\end{figure}
As seen from Fig. 1, the energy of the 2D polaron calculated by
taking into account the electron-phonon correlation within the
Buimistrov-Pekar approximation is lower than that of the 2D
adiabatic polaron. At $\alpha>\alpha_{c}\simeq2.94$, a bound state
of a carrier exists in 2D systems and the polaron remains localized
in the 2D polarization well.
\section{Formation of the intermediate-coupling bipolarons}
In the bipolaron problem the total energy of the interacting system
of two carriers and LO phonons is determined. The Hamiltonian and
wave function describing such a system are given by
\begin{eqnarray}\label{11Eq}
H_{B}=-\frac{\hbar^{2}}{2m^{*}}(\Delta_{1}+\Delta_{2})+
\frac{e^{2}}{\varepsilon_{\infty}r_{12}}+
\sum_{q}[V_{q}b_{q}(\exp(iqr_{1})+\exp(iqr_{2}))+\nonumber\\
V^{*}_{q}b^{*}_{q}(\exp(-iqr_{1})+\exp(-iqr_{2}))]+
\sum_{q}\hbar\omega_{o}b^{+}_{q}b_{q}
\end{eqnarray}
\begin{eqnarray}
\Psi=\Phi_{ph}\varphi(r_{1},r_{2})=U|0\rangle\varphi(r_{1},r_{2}),
\end{eqnarray}
where
\begin{eqnarray}
U=\exp\left[\sum_{q}\left(F_{q}(r_{1},r_{2})b^{+}_{q}-F^{*}_{q}(r_{1},r_{2})b_{q}\right)\right],
\end{eqnarray}
$\varphi(r_{1},r_{2})$ is the wave function of two carriers,
$V_{\vec{q}}$ is the form factor of the electron-phonon interaction
given by Eq. (\ref{10Eq}).

Averaging the Hamiltonian (\ref{11Eq}) over $\Phi_{ph}$, we obtain
the effective Hamiltonian
\begin{eqnarray}\label{12Eq}
\tilde{H}_{B}=-\frac{\hbar^{2}}{2m^{*}}(\Delta_{1}+\Delta_{2})+\frac{e^{2}}{\varepsilon_{\infty}r_{12}}+
\sum_{q}\left[\frac{\hbar^{2}}{2m^{*}}\left(|\nabla_{1}F_{q}(r_{1},r_{2})|^{2}+\right.\right.\nonumber\\
\left.\left.|\nabla_{2}F_{q}(r_{1},r_{2})|^{2}\right)+
\hbar\omega_{o}|F_{q}(r_{1},r_{2})|^{2}+
F_{q}(r_{1},r_{2})V_{q}(\exp(iqr_{1})+\right.\nonumber\\
\left.\exp(iqr_{2}))+F^{*}_{q}(r_{1},r_{2})V^{*}_{q}(\exp(-iqr_{1})+
\exp(-iqr_{2}))\right].
\end{eqnarray}
The displacement amplitude $F_{\vec{q}}(\vec{r}_{1},\vec{r}_{2})$ in
Eq. (\ref{12Eq}), which is a generalization of that applied above to
the polaron problem, can be written in the form
\begin{eqnarray}\label{13Eq}
F_{q}(r_{1},r_{2})=f_{q}+g_{q}(\exp(-iqr_{1})+\exp(-iqr_{2}))
\end{eqnarray}
Substituting Eq. (\ref{13Eq}) into Eq. (\ref{12Eq}), averaging the
Hamiltonian (\ref{12Eq}) over $\varphi(r_{1},r_{2})$, and minimizing
the energy $E_{B}=\langle\varphi|\tilde{H}_{B}|\varphi\rangle$ with
respect to $f_{\vec{q}}$ and $g_{\vec{q}}$, we obtain the following
functional for the total energy of a 2D bipolaron
\begin{eqnarray}\label{14Eq}
E_{B}=K_{1}+K_{2}+V_{12}-\sum_{q}\frac{\left|V_{q}\right|^{2}}{\hbar\omega_{o}}
W^{2}_{q}-\sum_{q}\frac{\left|V_{q}\right|^{2}}{\hbar\omega_{o}}
\frac{[D_{q}-W^{2}_{q}]^{2}}{D_{q}-W^{2}_{q}+2l^{2}_{0}q^{2}},
\end{eqnarray}
where
$$
K_{1,2}=\frac{\hbar^{2}}{2m^{*}}\int d^{2}r_{1}
d^{2}r_{2}\left[\frac{\partial\varphi(r_{1},r_{2})}{\partial
r_{1,2}}\right]^{2},\quad
V_{12}=\frac{e^{2}}{\varepsilon_{\infty}}\int\frac{\varphi^{2}(r_{1},r_{2})}{r_{12}}d^{2}r_{1}d^{2}r_{2},
$$
$$
W_{q}=\int
d^{2}r_{1}d^{2}r_{2}\left[\exp(iqr_{1})+\exp(iqr_{2})\right]\varphi^{2}(r_{1},r_{2})
$$
$$
D_{q}=\int d^{2}r_{1}d^{2}r_{2}[2+\exp(iq(r_{1}-r_{2}))+
\exp(-iq(r_{1}-r_{2}))]\varphi^{2}(r_{1},r_{2})
$$

The constant of the Coulomb interaction
$U=e^{2}/\varepsilon_{\infty}$ can be written as
$U=2\beta\sqrt{\alpha}$; where $\beta=1/(1-\eta)$,
$\eta=\varepsilon_{\infty}/\varepsilon_{0}$. The latter sum in Eq.
(\ref{14Eq}), just as in the polaron problem, is the correction term
arising from the electron-phonon correlation caused by the second
term in Eq. (\ref{13Eq}). The functional (\ref{14Eq}) without this
term $(g_{\vec{q}}=0)$ determines the energy of the strong-coupling
adiabatic bipolaron. For calculation of the bipolaron energy in the
strong-coupling limit, we choose the following two trial wave
functions:
\begin{eqnarray}\label{15Eq}
\varphi(r_{1},r_{2})=N\exp(-\delta^{2}(r^{2}_{1}+r_{2}^{2}))[1+\gamma\delta^{2}r_{12}^{2}],\quad
\gamma>0
\end{eqnarray}
\begin{eqnarray}\label{16Eq}
\varphi(r_{1},r_{2})=N\exp(-\delta^{2}(r^{2}_{1}+r_{2}^{2}))[1-b\exp(-\gamma\delta^{2}r_{12}^{2})],
\quad \gamma>0, \quad b<1
\end{eqnarray}
The bipolaron is stabile when the bipolaron energy is lower than
twice the polaron energy, i.e., the bipolaron stability regions is
determined by the inequality $E_{B}<2E_{p}$. At
$\alpha\rightarrow\infty$, the bipolaron energy calculated with the
use of the trial wave function (\ref{15Eq}) is approximately 2.5
times larger than the polaron energy $E_{p}=-(\pi/8)\alpha^{2}$,
i.e. $E_{B}/2E_{p}\approx1.25$. While the value of $E_{p}$
calculated in the strong-coupling limit ($\alpha\rightarrow\infty$)
with the use of the trial wave function (\ref{16Eq}) is equal to
$E_{B}\simeq2.58E_{p}$, so that $E_{B}/2E_{p}\simeq1.29$. In the
intermediate coupling regime, the energy of the 2D bipolaron is
calculated by using the trial wave function (\ref{16Eq}). The
results for the limit case $\beta=1$ ($\eta=0$) are shown in Fig.2.
\begin{figure}
\begin{center}
\includegraphics[width=0.6\textwidth]{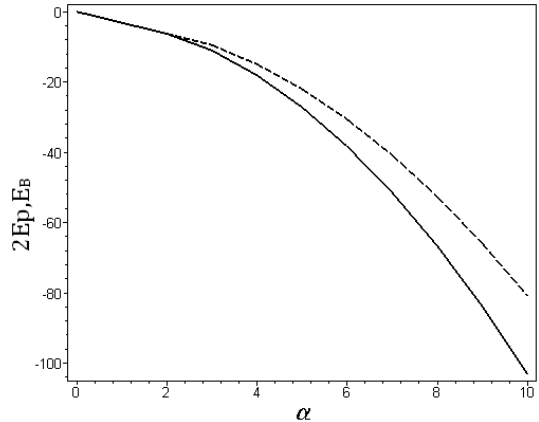}
\caption{\footnotesize The 2D bipolaron energy (solid curve),
calculated with using the Eq. (\ref{14Eq}), and the double 2D
polaron energy (dashed curve), calculated with using the Eq.
(\ref{5Eq}), as a function of $\alpha$ and in units of
$\hbar\omega_{o}$.}
\end{center}
\end{figure}
In this approximation the 2D bipolaron is stable for
$\alpha>\alpha_{c}=2.94$ which is comparable with the 2D results
obtained by using the Feynman path integral method \cite{9}. With
increasing $\alpha$, the bipolaron stability region is enlarged. For
example, at $\alpha=5$ and 10 the values of the parameter $\beta$
corresponding to the bipolaron formation region lie in the intervals
$1<\beta<1.201$ and $1<\beta<1.209$, respectively. For the cases
$\alpha=5$ and 10 the 2D bipolarons are stable at
$\eta<\eta_{c}=0.167$ and $\eta<\eta_{c}=0.173$, respectively.
\section{Conclusions}
In this work we have studied the formation of intermediate-coupling
optical polarons and bipolarons in 2D systems. The total energies of
the 2D polaron and bipolaron are calculated by using the
Buimistrov-Pekar method of canonical transformations. It is shown
that this method allows us to take into account the effects of the
electron-phonon correlation on the energies of the 2D (bi)polaron
for any values of the electron-phonon coupling constant $\alpha$ and
to derive the new expression for the polaron energy and limits for
strong and weak coupling. In the strong-coupling limit the
Buimistrov-Pekar method, the Feynman path integral method and the
LLP unitary transformation method give the same result for the
polaron energy. Our numerical results show that below a critical
value $\alpha_{c}$, the charge carriers are quasi-free (i.e.
delocalized) electrons or holes, and only at
$\alpha>\alpha_{c}\simeq2.94$ (bi)polarons exist in 2D polar
materials. The above critical values $\alpha_{c}$ and $\eta_{c}$ are
also close to the ones obtained by using the Feynman path integral
method.

Note that for calculation of the energies of 2D polaron and
bipolaron were used simple trial wave functions. Therefore, the
values of the (bi)polaron energies, $\alpha_{c}$ and $\eta_{c}$
obtained by these wave functions are approximate ones. For obtaining
more precise values of these parameters, it is necessary to use more
flexible wave functions, such as, for example, a sum of Gussians
\cite{24}.
\begin{center}
{\bf Acknowledgments}
\end{center}
We would like to thank B. Yavidov for his careful reading of the
manuscript. We are grateful to O.K. Ganiev for his technical
assistance. This work is supported by the Foundation of the
Fundamental Research of Uzbek Academy of Sciences grant FA-F2-F070.

\end{document}